\documentclass[cameraready]{Interspeech}

\title{SongBench: A Fine-Grained Multi-Aspect Benchmark for Song Quality Assessment}

\author[affiliation={1}, equalcontribution]{Dapeng}{Wu}
\author[affiliation={1}, equalcontribution]{Shun}{Lei}
\author[affiliation={2}]{Wei}{Tan}
\author[affiliation={2}]{Guangzheng}{Li}
\author[affiliation={2}]{Yunzhe}{Wang}
\author[affiliation={2}]{Huaicheng}{Zhang}
\author[affiliation={2}]{\protect\\ Lishi}{Zuo}
\author[affiliation={1}, correspondingauthor]{Zhiyong}{Wu}

\address{
    $^1$ Shenzhen International Graduate School, Tsinghua University, Shenzhen, China \\
    $^2$ Tencent
}

\email{\{wdp24, leis24\}@mails.tsinghua.edu.cn, zywu@se.cuhk.edu.hk}


\usepackage{comment}

\begin{document}

\maketitle

\begin{abstract}
Recent advancements in Text-to-Song generation have enabled realistic musical content production, yet existing evaluation benchmarks lack the professional granularity to capture multi-dimensional aesthetic nuances. In this paper, we propose \textbf{SongBench}, a specialized framework for fine-grained song assessment across seven key dimensions: Vocal, Instrument, Melody, Structure, Arrangement, Mixing, and Musicality. Utilizing this framework, we construct an expert-annotated database comprising 11,717 samples from state-of-the-art models, labeled by music professionals. Extensive experimental results demonstrate that SongBench achieves high correlation with expert ratings. By revealing fine-grained performance gaps in current state-of-the-art models, SongBench serves as a diagnostic benchmark to steer the development toward more professional and musically coherent song generation. The evaluation toolkit is available at \url{https://github.com/Tencent/SongBench}.
\end{abstract}

\section{Introduction}
The rapid advancement of generative Artificial Intelligence (AI) has transformed digital content creation, enabling the automated synthesis of complex multimedia content. Among these tasks, Text-to-Song generation is particularly challenging, as it requires high-fidelity audio synthesis, seamless integration of vocals and accompaniment, precise alignment between lyrical prosody and melody, and coherent long-term musical structure. Although recent representative models have achieved notable progress in generation quality \cite{yue, gong2025ace, diffrhythm, lei2025levo, yang2025songbloom}, the reliable evaluation of song generation remains a significant challenge.

Objective metrics such as Phone Error Rate, CLAP \cite{clap}, and Fréchet Audio Distance (FAD) \cite{fad} primarily measure controllability or distributional similarity, but struggle to reflect perceptual and artistic aspects of generated songs. Consequently, human subjective evaluation remains indispensable.
Compared to speech synthesis, where subjective quality evaluation has been systematically studied through Mean Opinion Score (MOS) prediction models \cite{speech0, speech1, speech2}, subjective assessment for song generation remains substantially more challenging due to its multidimensional attributes—such as melody, arrangement, and vocal performance—that require evaluators to possess musical literacy for nuanced differentiation. Moreover, the inherent variability in subjective preferences and personal experiences often leads to divergent judgments. This instability and lack of consistency restrict the fair comparison and optimization of models, ultimately hindering further advancements in the field.

Recent studies have begun to explore standardized pathways for music evaluation. MusicEval \cite{musiceval} pioneered an automated evaluation framework for instrumental music generation using an expert-annotated dataset, but its evaluation dimensions are restricted to overall quality and text consistency. In contrast, AudioBox \cite{AudioBox} attempted to build a universal system across speech, environmental sounds, and music, resulting in abstract metrics that lack specialized dimensions for musical artistry. Both frameworks primarily focus on evaluating short instrumental clips. As the first benchmark dedicated to the Text-to-Song task, SongEval \cite{songeval} introduced full-song evaluation by measuring subjective perceptions such as Coherence, Memorability, Naturalness, and Musicality. Despite this, SongEval faces two critical limitations. First, its evaluation dimensions exhibit significant semantic overlap, making it difficult for annotators to isolate inter-dimensional interference, which undermines evaluative objectivity. Second, as generative models rapidly iterate, basic qualities like coherence and naturalness have converged toward a high-performance plateau. This ``ceiling effect" is further exacerbated by a skewed distribution where scores disproportionately cluster in the highest brackets, causing severe rating compression. Consequently, there is an urgent need for a benchmark that offers superior aesthetic discernment, dimensional decoupling, and the granularity to distinguish between increasingly sophisticated AI-generated music.

\begin{figure}[t]
    \centering
    \includegraphics[width=\linewidth]{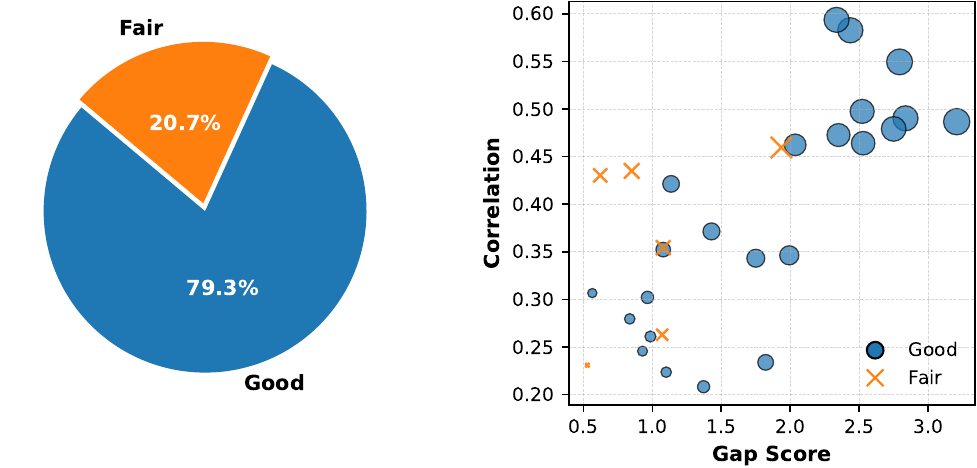}
    \caption{Expert candidate calibration. Proportion (left) and Performance distribution (right) of candidates.}
    \label{fig:expert_selection}
\end{figure}

To address these limitations, we propose a comprehensive, song-specific, and fine-grained evaluation framework for Text-to-Song generation. Rather than relying on ambiguous perceptual metrics, our framework is grounded in the essential elements considered during the actual song composition process. We define seven core dimensions: Vocal, Instrument, Melody, Structure, Arrangement, Mixing, and Musicality. This system achieves an organic integration of objective and subjective metrics, ensuring a comprehensive appraisal that spans from vocal to instrumental components and from local melody to global musical structures and holistic arrangements. Unlike perceptive dimensions prone to aesthetic bias, our framework decouples song evaluation into atomic metrics, each strictly focused on a single, concrete aspect of musical creation. This approach significantly reduces inter-dimensional coupling and provides more deterministic criteria for expert assessment, thereby substantially enhancing the reliability and consistency of the annotation results.

Based on this evaluation criteria, we recruited music experts to construct \textbf{SongBench}, a high-quality dataset comprising 11,717 samples generated by various open-source models and representative top-tier commercial systems. Through rigorous expert calibration, blind-listening, and quality filtering, SongBench ensures highly reliable and consistent annotations. To our knowledge, this is the largest benchmark for the Text-to-Song task to date, offering the broadest coverage and the most fine-grained evaluation.

Our contributions are summarized as follows: 1) Informed by the core elements of human musical composition, we establish a multi-dimensional, song-specific evaluation system, enabling robust and nuanced aesthetic assessment. 2) We construct the largest expert-annotated dataset with broad coverage and fine-grained evaluation dimensions, provides a solid foundation for developing and validating automated song assessment predictors. 3) Through extensive experiments, we demonstrate that SongBench achieves a high correlation with human expert ratings, providing a robust benchmark for song generation research.

\section{SongBench Dataset}

\subsection{Data Collection and Processing}
To construct a high-quality and diverse benchmark, we employed Hunyuan LLM \cite{hunyuan} to synthesize 4,000 lyrics and 384 prompts, which were randomly paired to generate diverse inputs. Based on these inputs, we collected 20,000 audio samples from multiple sources. Among them, 12,000 samples were generated using Suno \cite{suno2024} across versions v4 (6,500 samples), v4.5 (3,500 samples), and v5 (2,000 samples), reflecting the rapid evolution of generative models. We further integrated music from open-source architectures—comprising 3,000 samples each from LeVo \cite{lei2025levo} and SongBloom \cite{yang2025songbloom}, and 1,000 from ACE-Step \cite{gong2025ace}—along with 1,000 professionally produced copyrighted songs as high-quality human references, utilizing SongPrep \cite{songprep} for structure identification and lyric transcription. We will release all lyrics and prompts for reproducibility.

\begin{figure}[t]
  \centering
  \includegraphics[width=\linewidth]{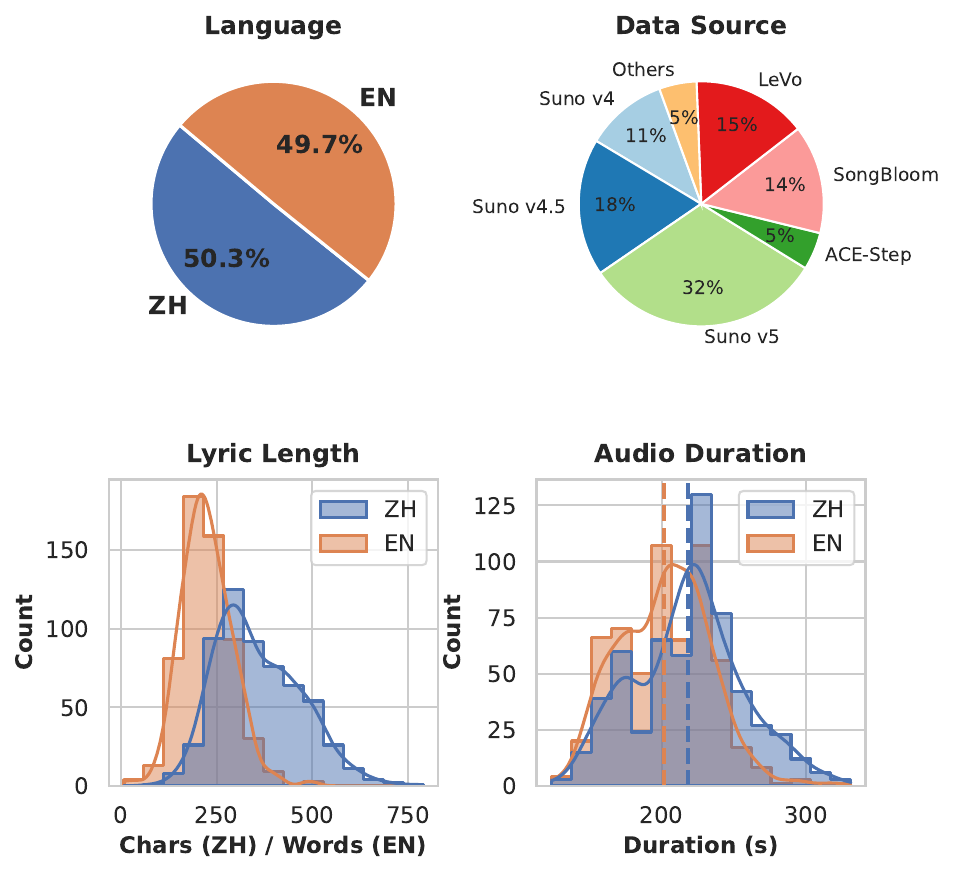}
  \caption{Statistical distributions of the SongBench dataset}
  \label{fig:data_dist}
\end{figure}

\begin{figure*}[t]
  \centering
  \includegraphics[width=0.62\linewidth]{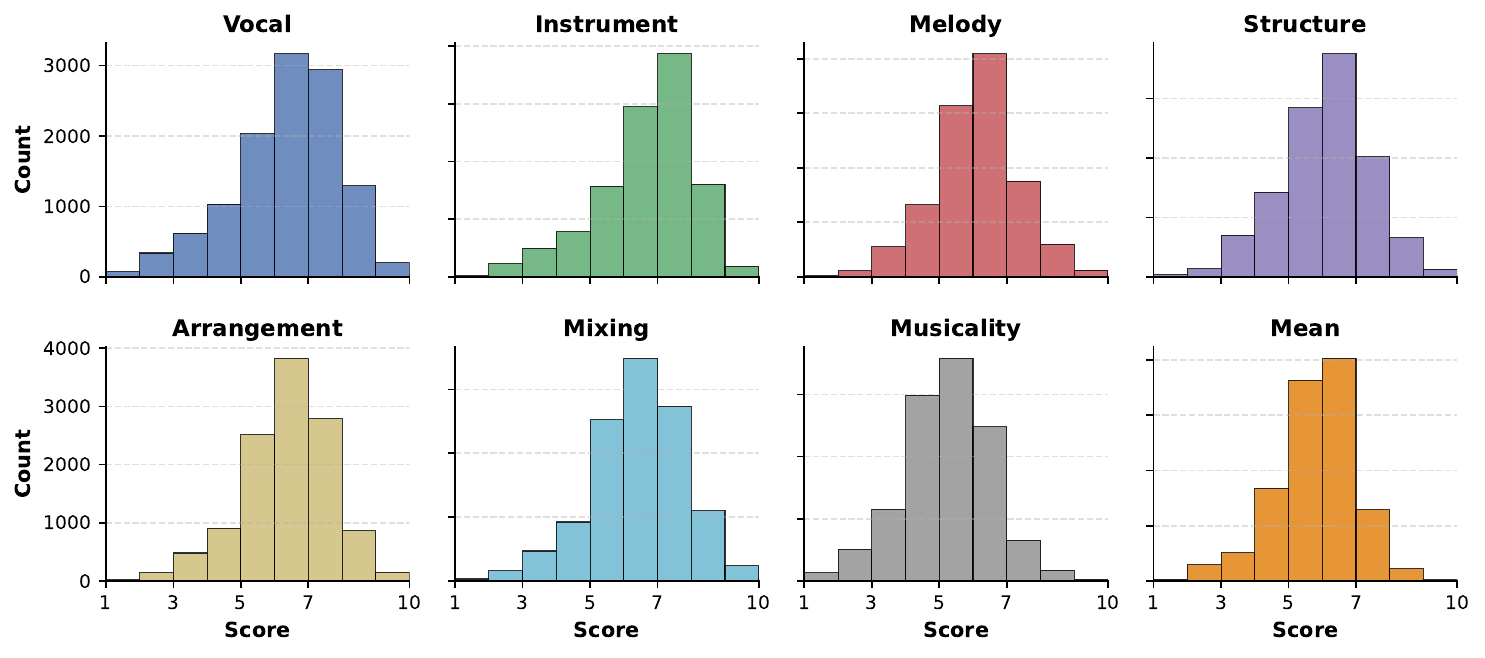}
  \caption{Score distributions across seven dimensions and the overall mean.}
  \label{fig:score_dist}
\end{figure*}

\subsection{Annotation Framework and Quality Control}
\subsubsection{Aesthetic Dimensions}

We propose a multi-dimensional evaluation framework where each dimension focuses on a single, specific element. The definitions of the seven dimensions are as follows:
\begin{itemize}
    \item \textbf{Vocal}: Assesses vocal quality, including clarity, pitch stability, and vocal techniques (e.g., vibrato, portamento).
    \item \textbf{Instrument}: Evaluates the realism and synthesis quality of instruments and their acoustic fidelity.
    \item \textbf{Melody}: Assesses melodic richness and memorability, focusing on the creative quality of the musical lines.
    \item \textbf{Structure}: Evaluates the organization of song sections (e.g., verse, chorus, bridge), ensuring natural transitions and adherence to compositional logic.
    \item \textbf{Arrangement}: Assesses the artistry of the harmonic framework and instrumental orchestration.
    \item \textbf{Mixing}: Evaluates post-production quality, focusing on the balance between tracks and the clarity of spatial imaging.
    \item \textbf{Musicality}: A holistic appraisal of the song’s overall artistic impact and auditory pleasure.
\end{itemize}

This framework evaluates song generation through a cohesive hierarchy that first addresses Vocal and Instrument quality to overcome the coarse limitations of metrics like FAD. Building on this, it scrutinizes the structural coherence and technical polish of Melody, Structure, Arrangement, and Mixing, ensuring a rigorous assessment of musical craftsmanship. Finally, by incorporating Musicality, the scheme captures overall artistic resonance and auditory pleasure. This integrated approach enables a precise diagnosis of an AI model’s performance, transitioning seamlessly from raw signal generation to high-level aesthetic expression.

\subsubsection{Expert Calibration}
To ensure professional annotation, we conducted a two-stage screening of 29 candidates with professional music backgrounds. As illustrated in Figure \ref{fig:expert_selection} (left), we first performed a qualitative ranking test. Candidates who failed to correctly identify the relative quality tiers of models were labeled ``Fair" and excluded. For the remaining ``Good" candidates, we further evaluated their professional sensitivity across two quantitative dimensions in Figure \ref{fig:expert_selection} (right): Pearson Correlation \cite{benesty2009pearson} (measuring rating alignment with the 4 elite experts) and Gap Score (reflecting the capacity to distinguish between models). Since an ideal expert should be both accurate and discerning, we selected the top 10 candidates clustered in the upper-right quadrant who demonstrated superior performance in both metrics. This rigorous process ensures our panel aligns precisely with high-level professional musical standards.

\subsubsection{Annotation Protocol and Data Filtering}
To ensure labeling consistency and mitigate subjective bias, we established a structured annotation protocol prior to large-scale deployment. Each song was evaluated by at least three music-specialized annotators across the seven dimensions using a 1–10 scale. To eliminate contextual interference, samples were presented in a randomized, double-blind manner, with their origins strictly concealed. To ensure high annotation quality, we monitored inter-annotator agreement and cross-referenced scores with performance priors to identify anomalies. Samples exhibiting excessive variance or ``straight-lining" patterns were flagged for secondary review or excluded. This rigorous filtering resulted in a final set of 11,717 high-quality samples, reflecting a stable and professional consensus on musical aesthetics.


\subsection{Data Distribution Statistics}
As shown in Figure \ref{fig:data_dist}, statistical analysis of the curated dataset (11,717 samples, $\sim$683.5 hours) reveals a balanced 1:1 ratio between Chinese and English tracks. With an average duration of 3.5 minutes, the dataset exhibits a normal distribution in lyric length, ensuring broad structural coverage. This refined collection is anchored by high-quality outputs from various Suno versions, followed by open-source models, and further enriched by authentic copyrighted songs, forming a diverse evaluation space that bridges synthetic and real-world music. Besides, Figure \ref{fig:score_dist} shows the expert scores across seven dimensions follow a normal distribution, effectively mitigating rating compression. This wide scoring gradient provides the resolution necessary to capture subtle aesthetic nuances, providing a solid empirical basis for developing robust automated predictors.

\section{Experimental Setup}

\subsection{Implementation Details}
We split the 11,717 high-quality samples into a training set and an In-Distribution (ID) test set with a 95:5 ratio. To assess generalization, we further construct an Out-of-Distribution (OOD) test set of 352 samples generated by external models, including DiffRhythm 2 \cite{jiang2025diffrhythm}, HeartMula \cite{yang2026heartmula}, ACE-Step v1.5 \cite{ace-step1.5}, MiniMax \cite{minimax}, Mureka \cite{mureka}, and Suno v4.5+ \cite{suno2024}, from 44 lyric–prompt pairs evenly covering mainstream musical styles. Following SongEval, we adopt the pre-trained MuQ \cite{zhu2025muq} as the self-supervised backbone to extract musical representations. Training is performed on 8 NVIDIA A100 GPUs with a batch size of 8, using AdamW and a CosineAnnealingLR scheduler with an initial learning rate of 1e-4.

\subsection{Evaluation Metrics}
Using SongEval as the baseline, we evaluate the alignment between model predictions and expert annotations at both the system and utterance levels. Evaluation metrics include Mean Absolute Error (MAE) to measure absolute prediction error, the Pearson Linear Correlation Coefficient (LCC) \cite{lcc} to quantify linear alignment with human ratings, and both Spearman’s Rank Correlation Coefficient (SRCC) \cite{srcc} and Kendall’s Tau (KTAU) \cite{ktau} to evaluate the consistency of relative quality ranking. 

\begin{table}[t]
\centering
\setlength{\tabcolsep}{3.5pt}
\caption{Correlation performance Across Different Dimensions. The ``\textemdash" denotes that MAE is not comparable due to SongEval's different scoring scale.}
\label{tab:main_results}
\begin{tabular}{lcccc}
\toprule
\multicolumn{5}{c}{\textbf{Utterance-level}} \\
\midrule
Dimensions & MAE $\downarrow$ & LCC $\uparrow$ & SRCC $\uparrow$ & KTAU $\uparrow$ \\
\midrule

Vocal        & 0.831 & 0.798 & 0.780 & 0.583 \\
Instrument   & 0.528 & 0.851 & 0.842 & 0.653 \\
Melody       & 0.895 & 0.781 & 0.786 & 0.591 \\
Structure    & 0.829 & 0.823 & 0.824 & 0.623 \\
Arrangement  & 0.790 & 0.850 & 0.851 & 0.657 \\
Mixing       & 0.824 & 0.844 & 0.834 & 0.628 \\
Musicality (Ours) & 0.738 & 0.835 & 0.829 & 0.636 \\
Musicality (SongEval) & \textemdash & 0.703 & 0.593 & 0.420 \\
\midrule
\multicolumn{5}{c}{\textbf{System-level}} \\
\midrule
Dimensions & MAE $\downarrow$ & LCC $\uparrow$ & SRCC $\uparrow$ & KTAU $\uparrow$ \\
\midrule

Vocal        & 0.774 & 0.961 & 0.929 & 0.810 \\
Instrument   & 0.378 & 0.990 & 0.964 & 0.905 \\
Melody       & 0.888 & 0.958 & 0.929 & 0.810 \\
Structure    & 0.814 & 0.950 & 0.893 & 0.714 \\
Arrangement  & 0.771 & 0.982 & 0.964 & 0.905 \\
Mixing       & 0.707 & 0.985 & 0.893 & 0.809 \\
Musicality (Ours)   & 0.713 & 0.976 & 0.929 & 0.810 \\
Musicality (SongEval) & \textemdash & 0.839 & 0.607 & 0.429 \\
\bottomrule
\end{tabular}
\end{table}

\section{Experimental results}

\begin{table*}[t]
\centering
\setlength{\tabcolsep}{3.5pt}
\caption{Model Performance Comparison Across Dimensions. For a direct comparison, SongEval’s five original dimensions are averaged to serve as its aggregate metric.}
\label{tab:model_full_comparison}
\begin{tabular}{lccccccccc}
\toprule
Model 
& Melody 
& Arrangement 
& Musicality 
& Vocal 
& Instrument 
& Mixing 
& Structure 
& Mean (Ours) 
& SongEval \\
\midrule

\multicolumn{10}{c}{\textit{Open-source Models}} \\
\midrule
Diffrhythm 2   & 5.16 & 5.09 & 4.15 & 5.24 & 5.10 & 4.72 & 5.02 & 4.93 & 3.71 \\
LeVo          & 5.10 & 5.23 & 4.33 & 5.56 & 5.87 & 5.18 & 5.08 & 5.18 & 3.49 \\
HeartMuLa     & 5.80 & 5.91 & 4.94 & 6.14 & 6.15 & 5.93 & 5.74 & 5.80 & 4.30 \\
ACE-Step v1.5   & 6.11 & 6.31 & 5.12 & 6.17 & 6.37 & 6.12 & 6.09 & 6.04 & 4.11 \\

\midrule
\multicolumn{10}{c}{\textit{Commercial Models}} \\
\midrule
MiniMax 2.0    & 6.37 & 6.61 & 5.39 & 6.80 & 6.67 & 6.39 & 6.33 & 6.37 & 4.21 \\
MiniMax 2.5    & 6.33 & 6.55 & 5.49 & 6.69 & 7.79 & 6.41 & 6.37 & 6.52 & 4.21 \\
Mureka O2      & 6.69 & 6.97 & 5.78 & 7.04 & 6.92 & 6.59 & 6.72 & 6.67 & 4.29 \\
Mureka V8      & 6.85 & 7.08 & 6.01 & 7.26 & 7.18 & 6.88 & 6.84 & 6.87 & 4.50 \\
Suno v4.5       & 6.60 & 6.80 & 5.72 & 6.94 & 6.83 & 6.56 & 6.72 & 6.60 & 4.45 \\
Suno v4.5+      & 6.81 & 7.04 & 5.81 & 7.18 & 7.10 & 6.82 & 6.80 & 6.79 & 4.45 \\
Suno v5         & 6.80 & 7.11 & 5.96 & 7.28 & 7.13 & 6.96 & 6.81 & 6.86 & 4.44 \\

\bottomrule
\end{tabular}
\end{table*}

\subsection{Correlation Analysis}
To evaluate system reliability, we conducted correlation analyses on the OOD test set at both utterance and system levels. As summarized in Table \ref{tab:main_results}, the results demonstrate strong alignment between the automated evaluation and expert human annotations.
\textbf{At the utterance level}, the model exhibits high predictive stability across all dimensions. LCC and SRCC for Melody, Arrangement, and Musicality consistently exceed 0.78. Notably, the low MAE scores across the board—particularly in the Instrument dimension (0.528)—reflect its precision in absolute score estimation. Furthermore, the KTAU values, which represent a stricter measure of ranking concordance, consistently remain above 0.58. This indicates that our model effectively preserves the relative quality hierarchy of individual tracks, even within the complex Vocal and Melody dimensions.
\textbf{At the system level}, the performance metrics surge significantly, highlighting the model’s efficacy in macro-trend assessment. All dimensions achieve LCCs above 0.95 and SRCCs between 0.89 and 0.96. The diminishing MAE at the system level suggests that the model’s bias is minimized when aggregating scores, while high KTAU values confirm its exceptional ability to rank different generative systems. Specifically, across both evaluation levels, our model consistently surpasses SongEval in the shared Musicality dimension, exhibiting significantly higher correlation and ranking consistency with human expert ratings.

\begin{table}[t]
\centering
\setlength{\tabcolsep}{6pt}
\caption{Pairwise Group-level Accuracy(\%) Comparison}
\label{tab:group_acc}
\begin{tabular}{lcc}
\toprule
Group & SongEval & Ours \\
\midrule
LeVo vs LeVo   & 43.48 & \textbf{64.13} \\
Suno vs Suno   & 55.10 & \textbf{62.24} \\
LeVo vs Suno   & 82.47 & \textbf{85.57} \\
\bottomrule
\end{tabular}
\end{table}

\subsection{Comparative Evaluation of Generative Models}
To evaluate the discriminative ability of our framework on high-aesthetic generation, we conducted a comparative analysis against SongEval, incorporating both ID and OOD test sets. As summarized in Table \ref{tab:model_full_comparison}, while both identify the performance gap between open-source and commercial models, our framework provides significantly higher evaluative resolution. SongEval exhibits a ``ceiling effect", assigning nearly stagnant scores to successive iterations of high-performing models—notably across the Suno (v4.5 to v5) and MiniMax (2.0 to 2.5) series. In contrast, our system successfully decodes their evolutionary trajectory, capturing incremental gains (e.g., Suno’s rise from 6.60 to 6.86 and MiniMax's leap to 6.52) that SongEval fails to differentiate. This capability is primarily driven by the normal distribution of our training labels, which avoids the rating compression and score saturation. By maintaining a fine-grained scoring gradient even among top-tier samples, our framework preserves the sensitivity needed to reward the refined musical nuances found in evolving commercial systems.

\subsection{AB Test}
To evaluate the discriminative granularity of our model, we conducted an AB test using 100 lyric–prompt pairs to generate 200 audio samples each from LeVo and Suno. We defined expert preferences as the ground truth, where a ``win" was recorded when one sample's score exceeded the other's by a specified threshold. Based on this, we compared our system's predictive accuracy against SongEval.
As summarized in Table \ref{tab:group_acc}, both systems perform reliably in inter-model comparisons, a relatively straightforward task due to the distinct quality gap between different generative systems. In this scenario, both achieve over 80\% accuracy, demonstrating effectiveness in distinguishing broad quality variances.
However, a clear performance gap emerges in the more challenging intra-model setting. On the LeVo task, SongEval’s accuracy drops sharply to 43.48\% and 55.10\% for Suno, both approaching near-random performance while our system maintains an accuracy exceeding 60\%. These results demonstrate that while SongEval suffers from perceptual blurring, our model exhibits superior fine-grained discernment, effectively capturing intricate aesthetic flaws overlooked by coarse-grained metrics.

\section{Conclusion}
In this paper, we introduce SongBench, a multi-dimensional evaluation system grounded in core musical elements to enable nuanced aesthetic assessment. We construct the largest expert-annotated dataset to date, featuring broad coverage and fine-grained evaluation dimensions to provide a solid foundation for training reliable assessment models. Extensive experiments demonstrate that SongBench aligns closely with human expert ratings and exhibits superior fine-grained aesthetic discernment, offering a robust and comprehensive benchmark for future research in song generation.

\newpage

\section{Generative AI Use Disclosure}
During the preparation of this manuscript, the authors used generative AI tools exclusively for the purpose of language editing and manuscript polishing to improve readability. These tools were not used to generate any core scientific ideas, experimental data, or technical contributions. All authors have thoroughly reviewed and approved the final version of the manuscript, and assume full responsibility for the integrity and entirety of its content.

\bibliographystyle{IEEEtran}
\bibliography{new_ref}

\end{document}